\documentclass[conference]{IEEEtran}
\IEEEoverridecommandlockouts
\usepackage[utf8x]{inputenc}
\usepackage{amsmath}
\usepackage{cite}
\usepackage{amsmath,amssymb,amsfonts}
\usepackage{algorithmic}
\usepackage{graphicx}
\usepackage{textcomp}
\usepackage{xcolor}

\def\BibTeX{{\rm B\kern-.05em{\sc i\kern-.025em b}\kern-.08em
    T\kern-.1667em\lower.7ex\hbox{E}\kern-.125emX}}
\begin{document}

\title{Cyberbullying in Text Content Detection: An  Analytical Review
}

\author{\IEEEauthorblockN{Sylvia W. Azumah, Nelly Elsayed, Zag ElSayed, Murat Ozer}
\IEEEauthorblockA{\textit{School of Information Technology} \\
\textit{University of Cincinnati}\\
Cincinnati, Ohio, United States \\
azumahsw@mail.uc.edu, elsayeny@ucmail.uc.edu, elsayezs@ucmail.uc.edu, ozermm@ucmail.uc.edu
}}


\maketitle
\thispagestyle{plain}
\pagestyle{plain}

\begin{abstract}
Technological advancements have resulted in an exponential increase in the use of online social networks (OSNs) worldwide. While online social networks provide a great communication medium, they also increase the user’s exposure to life-threatening situations such as suicide, eating disorder, cybercrime, compulsive behavior, anxiety, and depression.  
To tackle the issue of cyberbullying, most existing literature focuses on developing approaches to identifying factors and understanding the textual factors associated with cyberbullying. While most of these approaches have brought great success in cyberbullying research, data availability needed to develop model detection remains a challenge in the research space. This paper conducts a comprehensive literature review to provide an understanding of cyberbullying detection. 
\end{abstract}

\begin{IEEEkeywords}
Cyberbullying, Cybercrime, text detection, dataset, deep learning, online social network, social media. 
\end{IEEEkeywords}


\section{Introduction}
The use of social media as a technological device revolutionized the mode of communication and exchange of information in our current society. The ability to send information across thousands of miles at the speed of light has helped businesses in generating more revenue and allows Internet users to stay in contact with friends and family through various social media platforms. Social media has proven to be beneficial to humans by presenting a means of creating events, posting videos, and staying connected with loved ones while also negatively impacting the mental well-being of young individuals who are subject to cyberbullying on social media platforms~\cite{van2018automatic}. The Datareportal study found that 53\% of the global population uses social media platforms and that the average person spends about three hours a day on social media\cite{datareportal}. OSNs are used daily for a substantial amount of time. Cyberbullying has developed as a major problem on online social networks or media, afflicting both children and young adults \cite{lieberman2011let}. Generally, cyberbullying is identified as bullying that occurs via electronic devices or the Internet to cause harm or victimize someone~\cite{feinberg2009cyberbullying}. Cyberbullying has occurred and continues to occur on social networking sites such as Facebook, WhatsApp, Twitter, and Instagram. Reports of Soni et al.~\cite{soni2018see} indicate that 40\% of teenagers in the United States experienced cyberbullying harassment and name-calling, which negatively impacted their mental well-being, including deep emotional trauma, psychological and psychosomatic disarray~\cite{soni2018see}. According to Utemissova et al.~\cite{utemissova2021cyberbullying}, one-third of people bullied did not realize the negative behavior shown was associated with cyberbullying. Other findings from researchers stated that young adults who are cyberbullied have a higher tendency to harm themselves or have suicidal thoughts and behavior \cite{singh2017they}. Additionally, a recommendation in the paper Modeling the detection of textual cyberbullying by Dinakar et al.~\cite{dinakar2011modeling} highlights that about 83\% of young adults or people that are bullied believe that online social network platforms should have a part to play in curbing the issue of Cyberbullying on their social platforms~\cite{dinakar2011modeling}. 
In recent years, researchers have studied and understood most social science cyberbullying problems~\cite{nadali2013review}. Part of the preventive measures recommended by researchers to help with the cyberbullying problem include, but are not limited to, human interference, educational awareness and recommendation, deleting offensive terms, and cyberbullying detection~\cite{nadali2013review}. 

\begin{figure}
	\centering
	\includegraphics[width=0.55\linewidth]{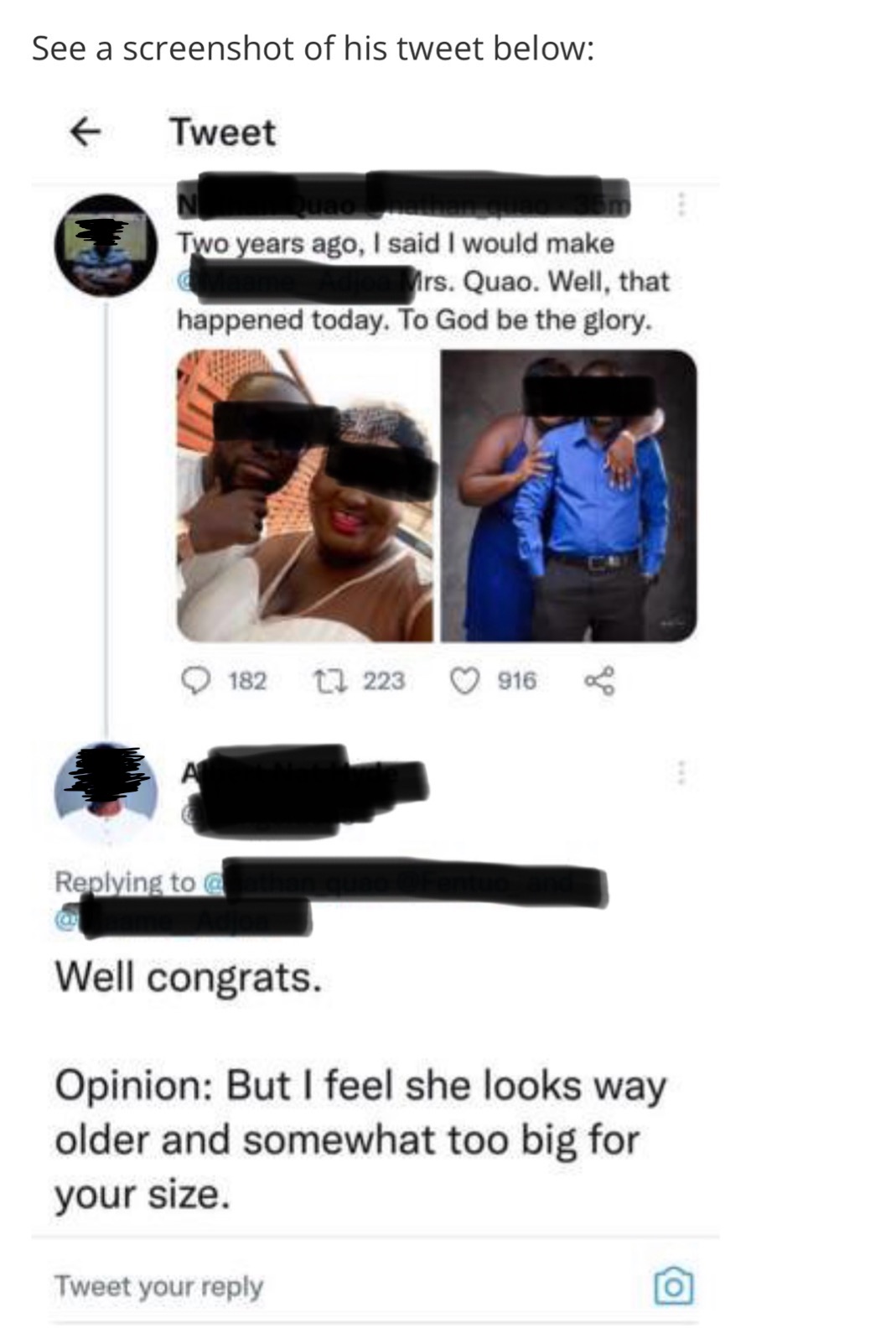}
	\caption{A real example of cyberbulling tweet captured from the social media site Twitter. Some parts of the image are blured to hide the identity of persons who are involved in the cyberbulling case~\cite{makafui_2022}.}
	\label{Twitter}
\end{figure}

Figure \ref{Twitter} depicts an example of cyberbullying on online social media networks or social media. They show demeaning words and phrases such as lo\$er, pi\$\$ off, sh*t, fat a\$\$, and other similar phrases to the victims. Over the years, most researchers have worked on solving the problem of cyberbullying. Grellety et al.~\cite{balakrishnan2019cyberbullying} developed queries in detecting cyberbullying contents with a precision of 91.25\%. Cyberbullying is often perceived as a defensive or aggressive response, with most perpetrators able to remain anonymous while victimizing targets. 
The research of Dalvi et al.~\cite{dalvi2020detecting} developed software for detecting cyberbullying posts on Twitter using machine learning algorithms. While most of these approaches have brought great success in cyberbullying research, data availability needed to develop model detection remains a challenge in the research space \cite{van2018automatic},~\cite{salawu2017approaches},~\cite{newman2020social}.
This paper provides a systematic literature review focusing on addressing the following research questions: 
\begin{itemize}
  \item {What are the major problems of cyberbullying detection for researchers?}
  \item {What are the impacts of law or legislation on cyberbullying across different continents? }
 \end{itemize}

This paper is structured as follows: 
Section~\ref{cyberbullying_forms} describes cyberbullying forms, content, crimes, and the policies pertaining to cyberbullying and its impact on an individual. Section~\ref{methodology} explains the method used to collect literature, the screening criteria used to identify the papers included in the review, and the criteria used to exclude literature not considered in the review.  
Afterward, Section~\ref{literature_review} discusses past, present, and future solutions in cyberbullying detection by first identifying cyberbullying, text detection, cyberbullying crime analysis, cyberbullying law, and cyberbullying data. Section \ref{analysis} presents the analysis and discussion of the study. 
Finally, Section~\ref{conclusion} provides an overview of the study via the research questions presented in this paper, followed by a recommendation for future research and the limitations of this study. 

\section{Cyberbullying Forms}\label{cyberbullying_forms}
As the Internet evolves, more cyberbullying occurs among its users. Over the years, researchers have come across various forms of Cyberbullying ranging from outing, stalking, and impersonation, just to mention a few. A recent investigation by Blumberg et al.~\cite{paul2012revisiting} introduces new forms of cyberbullying that have continuously evolved over the years. Some include trolling and doxing. 
Some of the significant cyberbullying forms are: 

\subsubsection{Flaming} 
In cyberbullying, the term flaming is described as an intense argument or disagreement via emails, instant messaging, and social media where the participants relay rude, offensive, profanity, threatening messages, and vulgar words \cite{feinberg2009cyberbullying}. For some social media users, flaming means to arouse or stir up anger and other negative emotions in a user\cite{alonzo2004flaming}. Other times, flaming can end up in a lawsuit when it becomes excessive and defamatory \cite{sandberg1994newsletter}. 

\subsubsection{Harassment}
According to Nandoli et al.~\cite{von2012cyberbullying}, harassment involves sending persistent insults and hurtful messages over the Internet.

\subsubsection{Stalking}
Cyberstalking is using the Internet or social media to consistently pursue, threaten and physically harm victims to the point where they start to look over their shoulder and feel unsafe~\cite{von2012cyberbullying}. Cyberbullies make use of stalking for two reasons, according to the research of Bocij et al.~\cite{bocij2002online}, i) To collect and gather personal identifiable information on the targeted person for further pursuit. ii) To consistently communicate with the target to threaten or inflict fear into them. 

\subsubsection{Denigration} 
It is to make available online defamatory information about their target. During denigration, cyberbullies post harmful materials such as text messages, videos, photos, gossip, and bank details of someone or a targeted person to humiliate them or damage their reputation \cite{bauman2015types}. 

\subsubsection{Impersonation} 
It is falsifying one’s identity to destroy their reputation or deceive people into hating them~\cite{aune2009cyberbullying}. 

\subsubsection{Outing} 
It is to make people's secrets or personal identifiable information available, such as health records and embarrassing information, online without the person's consent \cite{von2012cyberbullying}. 

\subsubsection{Trickery} 
As the name suggests, it is tricking or deceiving social media users to give out personal information about themselves, which is later shared without seeking their approval first \cite{aune2009cyberbullying}. 

\subsubsection{Trolling} 
In academia, trolling is defined in various forms. Some researchers identify trolling as behaving in a disruptive way on social media \cite{komacc2019overview}. On the other hand, some researchers carefully state that trolling has nothing to do with hostile intentions \cite{bishop2014representations}. Even though most researchers have not come to an agreement on defining what trolling is, most studies identify and agree that trolling is purposely done to harm others \cite{fichman2015bad}.  

\subsubsection{Exclusion}
Intentionally excludinterneting someone from an online social media group is termed as exclusion \cite{bishop2014representations}. For instance, leaving someone out of a gaming community. 

\subsubsection{Doxing}
It mainly occurs in social media platforms such as Twitter \cite{chen2019doxing}. Doxing is a form of cyberbullying that releases confidential or sensitive information without the person's consent. The doxed information includes but is not limited to demographic information, Social Security Number (SSN), ZipCode, Geographical Positioning System (GPS) coordinate, phone numbers, email address, passport number, username, and passwords \cite{chen2018doxing}, \cite{waseem2016hateful}. Disclosing these documents or sensitive information could lead to hate groups, child trafficking, spreading false information or rumors, and defamation \cite{karimi2022automated}. 

\subsubsection{Masquerading} 
Like impersonation, masquerading is a form of cyberbullying where cyber bullies adopt a false identity in order to harass or harm someone without revealing their true identity \cite{bauman2015types}. 

\subsubsection{Hate crime}
In cyberbullying hate crimes, the bully's behavior is often driven by their prejudice toward someone's race, religion, sexual orientation, gender, disability, or ethnicity~\cite{alnazzawi2022using}. 

\subsection{Cyberbullying Roles}
For cyberbullying to occur, it involves participants who accept well-structured roles\cite{van2018automatic}. 
Researchers in the field have identified various parties most prominent in cyberbullying. Vimala et al. \cite{balakrishnan2019cyberbullying} categorize cyberbullying roles into bullies or harassers, victims, bully-victims, and bystanders. Cyberbullies or Cyber-harassers are people who initiate bullying using an electronic device as a tool or the Internet; victims are the targeted person or the people that are bullied, and bystanders are witnesses to the bullying act online \cite{balakrishnan2019cyberbullying}. 

Further research by VanHee et al.~\cite{van2018automatic} highlights four cyberbullying roles as Harasser or bully, Victim, Bystander-defender, and Bystander-assistant. A bystander-defender is someone who protects and defends the victim from the bully, and a bystander-assistant is someone who encourages the bully or harasser to carry on with the act of cyberbullying. Even though traditional bullying studies have focused on just two bullying roles which are victim and bullies \cite{salmivalli1996bullying}, they acknowledge the importance of having a bystander as part of a bullying incident \cite{salmivalli2010bullying}, \cite{bastiaensens2014cyberbullying}.

\section{Methodology}\label{methodology}
This section presents a systematic literature review method used to collect and review the literature for this study. This research method is guided by the research questions highlighted earlier in the introduction of the study. Cyberbullying text detection literature that has been done between the years 2017 to 2022 is studied to identify the research gap proposed. To accomplish the systematic research review, the Jorgensen survey methodology (DICARe) is used \cite{amiruddin2019systematic}. The five stages of the Jorgensen method are explained as used below: 
\subsubsection{Define}The purpose and scope of the research are defined using the research questions: What are the major problems of cyberbullying detection for researchers? In addition, what are the impacts of law or legislation on cyberbullying across different continents? As stated in the introduction. The topic selected was cyberbullying text detection; the period of the study was performed over the last five years, from 2017 to 2022. Given the increase in cyberbullying over the years, the period from 2017 to 2022 looks appropriate for the study. 
\subsubsection{Identification}This phase identifies research papers that are relevant to the topic of interest. These papers were written in English and had abstracts, titles, research topic keywords, and conclusions.  
\subsubsection{Classify}The classification stage involves putting together research papers related to the research on a particular approach. For this systematic literature review, we used Cyberbullying detection as the main theme. We provided grouping based on text detection, the impact of cyberbullying, cyberbullying crime analysis, and recommendations for cyberbullying. 
\subsubsection{Analyze}This phase concentrates on classification in the previous step and analyzes the grouping to ensure it is correct and can be selected for further analyzing. 
\subsubsection{Report}This phase concentrates on classification in the previous step and analyzes the grouping to ensure it is correct and can be selected for further analysis. 
\subsection{Research Keywords}
To conduct the research methodology, the keywords were extracted from the research questions. 
The following keywords are: 
“Cyberbullying,” “Text or textual,” “Detection,” “Social media or Online social network.”

\subsection{Study Selection}
Publications must meet the following criteria to be considered for inclusion:
\begin{enumerate}
    \item The last five years (2017-2022)
    \item Cyberbullying
    \item Cyberbullying text detection
    \item Implications of cyberbullying
     \item Automatic cyberbullying detection
      \item Cyberbullying on social media or online social networks
\end{enumerate}
 
The following publications were excluded from the study selection: 
\begin{enumerate}
    \item Non-English papers
    \item Extended abstracts
    \item Cyberbullying image detection
    \item Poster presentations
     \item Short papers
      \item Books
      \item Survey
\end{enumerate}

\subsection{Search and selection strategy}
The search and selection strategy is a combination of different types of categories to extensively search for cyberbullying text detection papers. The goal of the search is to find relevant research papers in cyberbullying detection; 
\subsubsection{ Database Search} Using online inquiry components of popular publication databases is the most notable approach to scan for papers\cite{rajmohan2022decade}. This paper used some of the most popular publication databases, namely IEEE Xplore, Scopus, Web of Science, and ACM, to search for the papers related to the study. However, only papers from IEEE Xplore and Web of Science were used. This was used 
as these libraries provide sufficient keyword paper indexing that helps to find the most-related publications. The keywords that have been used in the publication's search are represented in Table \ref{tab:my-table1}.

\begin{table*}[]
\caption{Research database search and results.} 
\label{tab:my-table1}

 \renewcommand*{\arraystretch}{2}
\resizebox{\textwidth}{!}
{%
\begin{tabular}{|l|l|c|c|}
\hline
\textbf{Database}                                             & \textbf{Search Keyword  }                                                           & \textbf{Results}                             & \textbf{Filter}                                        \\ \hline
\multicolumn{1}{|c|}{IEEE Xplore}             & \multicolumn{1}{c|}{(“All metadata”: cyberbullying) AND (“All metadata”: “text”) AND (“All metadata”: “detection”) AND (“All metadata”: “system”)}                                         & \multicolumn{1}{c|}{21}        & \multicolumn{1}{c|}{2017-2022}  \\ \hline
\multicolumn{1}{|c|}{Web of Science}             & \multicolumn{1}{c|}{(((((ALL=(cyberbullying)) AND ALL=(detection)) AND 
ALL=(text))) AND ALL=(system)}                                         & \multicolumn{1}{c|}{34}        & \multicolumn{1}{c|}{2017-2022}  \\ \hline
\end{tabular}
}
\end{table*}

After every search, the results of the selected papers are reviewed by going through the abstract, title, and skimming through the contents of the paper. After that, duplicated copies found are taken out. Eventually, the results from each database search are merged. A total of 55 papers were returned from the search query. However, after skimming through to select the relevant papers for the study, 31 papers were selected and used. 

\section{Related Literature}\label{literature_review}

Cyberbullying or electronic bullying is one of the well-known risks of technological evolution and consists of voluntary and repeated actions against one or more individuals using computers and electronic devices \cite{aboujaoude2015cyberbullying}. In the past, traditional cyberbullying has been restricted to schools and the youth. However, with the rapid evolution of the Internet and the use of social media, cyberbullying has developed as a pressing issue beginning at home. Studies in this area show the increased rate of cyberbullying over the years~\cite{yusuf2019cyberbullying}.  
Most cyberbullying research efforts have focused on areas including cyberbullying online detection through text and images. Others have worked on providing recommendations to concerned parties such as Internet users, social media companies, law enforcement agencies, and many others with the intention of creating awareness on the topic. However, a common issue surrounding the study is the lack of a publicly available dataset for research analysis \cite{van2018automatic}, \cite{salawu2017approaches}, \cite{newman2020social}. 
This section examines literature in the domain of cyberbullying detection between 2017 and 2022. 

\subsection{Cyberbullying Text Detection }
Over the years, research done in cyberbullying text detection has constituted using both machine learning and filtration software \cite{haidar2017multilingual}. In social media platforms, filtration software is used to automatically filter unhallowed words or content \cite{nahar2013effective}. Other works applied machine learning and deep learning techniques for cyberbullying text detection rather than filtration techniques. With machine learning and deep learning techniques, comments or text are scrapped from social media platforms such as Twitter, Facebook, Instagram, and vine, to mention a few. These collected comments are then introduced to the machine learning and deep learning algorithms responsible for classifying cyberbullying.

Haidar et al.~\cite{haidar2017multilingual}, in their paper, detected cyberbullying in Arabic content. The researchers discovered that more research had been done on detecting cyberbullying in English content; hence they saw the need to explore detecting Arabic content. To solve this issue of cyberbullying, Haider et al.~\cite{haidar2017multilingual} proposed a machine learning algorithm that splits the dataset into training and testing data for classification. They also used the Dataiku DSS and WEKA to test the machine learning algorithm in their research. According to the researchers, the WEKA toolkit was used because it aids in the visualization, preparation, and classification of the Arabic language \cite{haidar2017multilingual}. The results from this study showed promising results from the use of the Support Vector Machine (SVM) and Naive Bayesian algorithm to detect cyberbullying. These two machine learning algorithms were used in this study based on past research work in cyberbullying.

Moreover, researchers have concluded that SVM and Naive Bayesian algorithms are the best algorithms for detecting cyberbullying text \cite{mclachlan2004discriminant}. Their module obtained an F1-Score of 92\% and 90\% consequently. 
Similar to \cite{haidar2017multilingual}, Rohit et al.\cite{pawar2019multilingual} detected cyberbullying behavior in Hindi and Marathi social media text content. The algorithms selected by Rohit et al. are Multinomial Naive Bayes (MNB), Logistics Regression (LR), and Stochastics Gradient Descent (SDG) to perform ten-fold cross-validation experimentation for the study \cite{pawar2019multilingual}. The algorithms selected in this study were selected based on their performance in previous studies \cite{pawar2019multilingual}. The results of their experiments showed 97\% accuracy and 96\% for F1-Score. According to the researchers, the results of their studies performed very well in detecting cyberbullying in both languages. In this way, their model can therefore detect multiple languages in the cyberbullying space
\cite{pawar2019multilingual, thangiah2012framework}.

Automatic detection of cyberbullying in social media text by Van Hee et al. \cite{van2018automatic} aimed at exploring cyberbullying automatic detection on social media. This automatic detection included discovering cyberbullying signals, such as the various forms of cyberbullying, and identifying posts from bullies, victims, and bystanders\cite{van2018automatic}. This study leveraged a binary classification to detect cyberbullying in English and Dutch. In addition, the paper highlighted the issue of few publicly available datasets, making it difficult for researchers to conduct constructive research addressing cyberbullying. To conclude, the classification revealed that the binary approach is promising to detect cyberbullying automatically in the future. However, a dive into using deep learning techniques to detect cyberbullying automatically would improve the ability of the classifier to predict cyberbullying content.
In ~\cite{lu2020cyberbullying}, Lu et al. proposed a Character-level Convolutional Neural Network with shortcuts (Char-CNNS) for the detection of cyberbullying in a social media text. Additionally, as part of trying to solve the issue of publicly available data for cyberbullying detection mentioned in \cite{van2018automatic},\cite{balakrishnan2019cyberbullying},\cite{dalvi2020detecting},\cite{salawu2017approaches}, \cite{newman2020social}, \cite{lu2020cyberbullying}, they contributed to making their dataset available on GitHub by crawling data from Sina Weibo comment\cite{lu2020cyberbullying}. The model had a precision, accuracy, and F1-score of 79\%, 71.6\%, and 69.8\%, respectively. The researchers concluded that the results of the performance of the Char-CNNS were competitive and outperformed various state-of-the-art algorithms that have been used in cyberbullying detection when compared. 
While most researchers have leveraged machine learning techniques in detecting cyberbullying text in social media, others are leaning towards using deep learning algorithms. A tremendous work by Iwendi Celestine et al.\cite{iwendi2020cyberbullying} developed a solution-based deep learning architecture to detect cyberbullying on social media. The study collected data from Facebook, Twitter, and Instagram preprocessed it and experimented on it using four deep learning algorithms, namely, Bidirectional Long Short-Term Memory (BLSTM), Gated Recurrent Units (GRU), Long Short-Term Memory (LSTM), and finally Recurrent neural network (RNN). In evaluating the performance of the deep learning algorithms, it was observed that the BLSTM model outperformed RNN, LSTM, and GRU with higher accuracy and F1-Score. Iwendi Celestine et al. concluded that their most performed deep learning algorithm would allow future researchers to leverage to solve cyberbullying. 

Fang, Yong et al.~\cite{fang2021cyberbullying}, conducted an experiment using three datasets of which two were collected from Twitter and the other from Wikipedia. The data was fed into a combination of a bidirectional GRU and a self-attention algorithm to detect textual cyberbullying content. It was observed in the study that the proposed methods performed tremendously high in all performance metrics. Table \ref{tab:my-table} is a representation of studies, metrics, approach categories, and classification algorithms used in cyberbullying detection. 
The studies highlighted in the table are ordered based on the year of publication, 2017 to 2022, as mentioned in the inclusion criteria.

\begin{table*}[]
\caption{Summarization of cyberbullying detection study. Information that has not been provided by the authors is indicated by the symbol (-).} 
\label{tab:my-table}

 \renewcommand*{\arraystretch}{2}
\resizebox{\textwidth}{!}
{%
\begin{tabular}{|l|c|c|c|c|}
\hline
\textbf{Study}                                               & \textbf{Year}                                                              & \textbf{Classifiers}                              & \textbf{Metrics}                                 & \textbf{Category}                               \\ \hline
\multicolumn{1}{|l|}{Haidar et al.~\cite{haidar2017multilingual}}             & \multicolumn{1}{c|}{2017}                                         & \multicolumn{1}{c|}{SVM, Naive Bayesian}        & \multicolumn{1}{c|}{F1-Score} & \multicolumn{1}{c|}{Machine Learning}  \\ \hline

\multicolumn{1}{|l|}{Alduailej et al.~\cite{alduailej2017challenge}}          & \multicolumn{1}{c|}{2017}         & \multicolumn{1}{c|}{Text Mining, Lexicon}                 & \multicolumn{1}{c|}{-}             & \multicolumn{1}{c|}{-}                  \\ \hline

\multicolumn{1}{|l|}{Van Hee et al.~\cite{van2018automatic}}             & \multicolumn{1}{c|}{2018}                                         & \multicolumn{1}{c|}{Binary Classification}        & \multicolumn{1}{c|}{Accuracy, Precision, Recall, F1-Score} & \multicolumn{1}{c|}{Machine Learning}  \\ \hline

\multicolumn{1}{|l|}{Shane et al.~\cite{murnion2018machine}}              & \multicolumn{1}{c|}{2018}                    & \multicolumn{1}{c|}{MAN, SAC}                & \multicolumn{1}{c|}{-}    & \multicolumn{1}{c|}{Sentiment Analysis}                  \\ \hline

\multicolumn{1}{|l|}{Lu et al.~\cite{lu2020cyberbullying}}             & \multicolumn{1}{c|}{2019}                                         & \multicolumn{1}{c|}{Char-CNN}        & \multicolumn{1}{c|}{Accuracy, Precision, F1-Score} & \multicolumn{1}{c|}{Deep  Learning}  \\ \hline
 
\multicolumn{1}{|l|}{Rohit et al.~\cite{pawar2019multilingual}}             & \multicolumn{1}{c|}{2019}                                         & \multicolumn{1}{c|}{MNB, LR, SDG}        & \multicolumn{1}{c|}{Accuracy, F1-Score} & \multicolumn{1}{c|}{Machine Learning}  \\ \hline

\multicolumn{1}{|l|}{Mahlangu et al.~\cite{mahlangu2019deep}}          & \multicolumn{1}{c|}{2019} & \multicolumn{1}{c|}{LSTM + Stacked Embedding}                & \multicolumn{1}{c|}{Accuracy, F1-Score}    & \multicolumn{1}{c|}{Deep Learning}                  \\ \hline

\multicolumn{1}{|l|}{Pascucci et al.~\cite{pascucci2019computational}}           & \multicolumn{1}{c|}{2019}                       & \multicolumn{1}{c|}{RF, Sequential Optimization Algorithm}                & \multicolumn{1}{c|}{Precision, Recall, F1-Score}    & \multicolumn{1}{c|}{Machine Learning}                  \\ \hline

\multicolumn{1}{|l|}{Andleeb et al.~\cite{andleeb2019identification}}            & \multicolumn{1}{c|}{2019}                        & \multicolumn{1}{c|}{SVM, Bernoulli NB}                & \multicolumn{1}{c|}{Accuracy, Precision, Recall, F1-Score}    & \multicolumn{1}{c|}{Machine, Deep Learning}                  \\ \hline

\multicolumn{1}{|l|}{Iwendi Celestine, et al.\cite{iwendi2020cyberbullying}}             & \multicolumn{1}{c|}{2020}                                         & \multicolumn{1}{c|}{BLSTM, GRU, LSTM, RNN}        & \multicolumn{1}{c|}{Accuracy, Precision, Recall,  F1-Score} & \multicolumn{1}{c|}{Deep  Learning}  \\ \hline

\multicolumn{1}{|l|}{Muneer et al.~\cite{muneer2020comparative}}             & \multicolumn{1}{c|}{2020}             & \multicolumn{1}{c|}{LR, SVM, RF, NB, LGBM}                & \multicolumn{1}{c|}{Accuracy, Precision, Recall, F1-Score}    & \multicolumn{1}{c|}{Machine Learning}                  \\ \hline

\multicolumn{1}{|l|}{Van B. et al.~\cite{van2020multi}}             & \multicolumn{1}{c|}{2020}                & \multicolumn{1}{c|}{SVM, CNN, XGBoost}                & \multicolumn{1}{c|}{Accuracy, ROC, F1-Score}    & \multicolumn{1}{c|}{Machine, Deep Learning}                  \\ \hline

\multicolumn{1}{|l|}{Wang et al.~\cite{wang2020multi}}               & \multicolumn{1}{c|}{2020}                       & \multicolumn{1}{c|}{BLSTM, MLP}                & \multicolumn{1}{c|}{Accuracy, F1-Score}    & \multicolumn{1}{c|}{Deep Learning}                  \\ \hline

\multicolumn{1}{|l|}{Talpur et al.~\cite{talpur2020cyberbullying}}          & \multicolumn{1}{c|}{2020}         & \multicolumn{1}{c|}{NB, KNN, RF, SVM, DT}                 & \multicolumn{1}{c|}{AUC, Precision, Recall, F1-Score}             & \multicolumn{1}{c|}{Machine, Deep Learning}     \\ \hline

\multicolumn{1}{|l|}{Fang, Yong et al.~\cite{fang2021cyberbullying}}        & \multicolumn{1}{c|}{2021}                   & \multicolumn{1}{c|}{BLSTM + Self-attention}                & \multicolumn{1}{c|}{Precision, Recall, F1-Score}    & \multicolumn{1}{c|}{ Deep Learning}                  \\ \hline

\multicolumn{1}{|l|}{Parera et al.~\cite{perera2021accurate}}            & \multicolumn{1}{c|}{2021}           & \multicolumn{1}{c|}{SVM, LR}                & \multicolumn{1}{c|}{Accuracy, Precision, Recall, F1-Score}    & \multicolumn{1}{c|}{Machine Learning}                  \\ \hline

\multicolumn{1}{|l|}{Kumar et al.~\cite{kumar2021multimodal}}              & \multicolumn{1}{c|}{2021}         & \multicolumn{1}{c|}{CNN}                & \multicolumn{1}{c|}{Accuracy, Precision, Recall}    & \multicolumn{1}{c|}{Deep Learning}                  \\ \hline

\multicolumn{1}{|l|}{Perera et al.~\cite{perera2021accurate}}              & \multicolumn{1}{c|}{2021}         & \multicolumn{1}{c|}{SVM, LR}                & \multicolumn{1}{c|}{Accuracy, Precision, Recall, F1-Score}    & \multicolumn{1}{c|}{Machine Learning}                  \\ \hline

\multicolumn{1}{|l|}{Yuvaraj et al.~\cite{yuvaraj2021automatic}}              & \multicolumn{1}{c|}{2021}         & \multicolumn{1}{c|}{DNN}                & \multicolumn{1}{c|}{Accuracy, G-Mean, F1-Score}    & \multicolumn{1}{c|}{Deep Learning}                  \\ \hline

\multicolumn{1}{|l|}{Bozyiugit et al.~\cite{bozyiugit2021cyberbullying}}              & \multicolumn{1}{c|}{2021}         & \multicolumn{1}{c|}{SVM, AdaBoost, RF, LR}                & \multicolumn{1}{c|}{Accuracy, Precision, Recall, F1-Score}    & \multicolumn{1}{c|}{Machine Learning}                  \\ \hline

\multicolumn{1}{|l|}{kazi et al.~\cite{alam2021cyberbullying}}              & \multicolumn{1}{c|}{2021}         & \multicolumn{1}{c|}{XGBoost, LR}                & \multicolumn{1}{c|}{Accuracy, AUC,  F1-Score}    & \multicolumn{1}{c|}{Machine Learning}                  \\ \hline

\multicolumn{1}{|l|}{Jahan et al.~\cite{jahan2022data}}              & \multicolumn{1}{c|}{2022}         & \multicolumn{1}{c|}{CNN, FastText, BERT}                & \multicolumn{1}{c|}{Accuracy, F1-Score}    & \multicolumn{1}{c|}{Deep Learning}                  \\ \hline

\multicolumn{1}{|l|}{Singh et al.~\cite{singh2022deep}}              & \multicolumn{1}{c|}{2022}         & \multicolumn{1}{c|}{CNN, BERT}                & \multicolumn{1}{c|}{Accuracy, Precision, Recall, F1-Score}    & \multicolumn{1}{c|}{Deep Learning}                  \\ \hline

\multicolumn{1}{|l|}{Kumar et al.~\cite{kumar2022bi}}              & \multicolumn{1}{c|}{2022}         & \multicolumn{1}{c|}{Bi-GRU Attention-CapsNet}                & \multicolumn{1}{c|}{ROC-AUC, F1-Score}    & \multicolumn{1}{c|}{Deep Learning}                  \\ \hline

\end{tabular}
}
\end{table*}

\subsection{Cybercrime analysis on cyberbullying}
A cybercrime is using the Internet as a tool or any electronic device to commit a crime. Cybercrimes include but are not limited to child pornography, cyberbullying, harassment, doxing, hate crime, impersonation, and many others\cite{andleeb2019identification}. The United States Department of Justice amplified the definition of cybercrime as an illegal use of an electronic device or computer to store evidence~\cite{nadali2013review}. Thangiah et al.\cite{thangiah2012framework} categorize cybercrime into two, namely, content-based and technology-based crimes. 
 Over the years, research in hate crime or hate speech has increased \cite{elsafoury2021timeline}. The main aim of this paper is to review cyberbullying text detection literature in the past five years. However, hate crime or speech is added to the literature due to its similarities with cyberbullying detection. In a paper on detecting and visualizing hate speech in social media, Modha et al.\cite{modha2020detecting} studied hate crimes on Twitter and Facebook by exploring use case scenarios. They developed a visualization tool to find and display hate crime content posted by online social media users. They observed that while most researchers have leveraged the use of machine learning algorithms to detect text in hate speech or crime, they have also presented a learning process drawback \cite{modha2020detecting}. They went further to highlight that deep learning algorithms will not only perform better in hate speech text detection but has the tendency to solve the machine learning limitations. In a framework for hate speech detection using deep convolutional neural networks by Roy et al.\cite{roy2020framework}, they develop an automated deep learning system together with GloVe embedding vector to capture hate speech text from Twitter. The outcome of the experimentation showed a poor accuracy rate on the train test dataset. However, with 10-fold cross-validation used with the deep learning algorithm, the model achieved a good prediction recall value. Alnazzawi et al.\cite{salawu2017approaches} proposed a method to develop hateMotiv corpus to provide publicly available hate crime data. The hate crime data included the types of hate crimes and the motives behind committing the crime. Additionally, the authors worked on a vocabulary list that can be used as a resource for hate crimes and their motivation. 
\section{Analysis and Discussion}\label{analysis}
\subsection{Major problems of cyberbullying detection for researchers}
\subsubsection{Publicly available data}
After analyzing the selected papers for the literature review, it was noticed that most researchers mentioned that having a publicly available dataset for the classification of cyberbullying text was and is still a challenge in cyberbullying. For instance, Salawu et al.\cite{salawu2017approaches} approach to automated detection of cyberbullying paper identified that most studies used online harassment, insult, and hate crime datasets for cyberbullying detection. These kinds of data consist of individual abusive or insulting content, are outdated, and as such, are inappropriate for creating and selecting features for cyberbullying detection but rather beneficial for detecting a particular form of cyberbullying such as cyber aggression, hate crime, and harassment.
\subsubsection{Heterogeneity of data}
Cyberbullying is not restricted to English alone but also extends to other languages, such as but not limited to Arabic, Turkish, Dutch, and Hindi. Cyberbullying detection research has leveraged different language data to solve cyberbullying on online social media. Due to language diversity, text classification in cyberbullying has become challenging as the translation of the text can be ambiguous. Additionally, it becomes difficult for researchers to understand the dataset\cite{lu2020cyberbullying}. 
As per their recommendation of how to solve the unavailability of cyberbullying dataset highlighted as a challenge in these studies \cite{van2018automatic},\cite{balakrishnan2019cyberbullying}, \cite{dalvi2020detecting,salawu2017approaches,fang2021cyberbullying}, they urge researchers in the field to work on developing dataset that could go a long way to helping the cyberbullying research community detect and recognize the various forms of cyberbullying, cyberbullying roles and mapping the relations between the victim and offenders in cyberbullying. So far, the most popular cyberbullying dataset is the Trace dataset \cite{salawu2017approaches}. Although it consists of cyberbullying and non-cyberbullying tweets and some forms of cyberbullying, researchers are still encouraged to put much effort into creating datasets so that machine learning and deep learning models can perform advanced cyberbullying detection tasks \cite{salawu2017approaches}.   

With cyberbullying being an issue of great importance to society, it is important to pay attention to cyberbullying prevention and improvement methods. These are only feasible if parents, law enforcement, social media platforms, educational colleges or educators, and researchers make a conscious effort to create awareness of the topic through knowledge dissemination. It is only when this occurs that cyberbullying detection research can progress worldwide. 

\subsection{Impact of law or legislation on cyberbullying across different continents.}
While it is important that law enforcement agencies take a forward-looking approach to combating cyberbullying, the law controlling the use of electronic devices and social media has fallen behind. Hence, making it difficult for law enforcement to handle cyberbullying cases\cite{law}. 
Through research, it is noted that most advanced countries such as Canada, America and 
Australia has adopted legal ways of governing cyberbullying \cite{tagaymuratovna2022cyberbullying}. In the United States of America, the law requires that differentiation is constructed around Cyberbullying, Electronic harassment, and Cyberstalking for legal rules to be applied. 

According to research in~\cite{law,tagaymuratovna2022cyberbullying}, about forty-nine American states have traditional bullying laws, forty states have laws pointing to cyber or electronic harassment, and finally, fourteen states have laws governing cyberbullying. For instance, in New Jersey, a revenge porn law has been established to prevent the leak of nude photos and videos of minors without their consent\cite{matwyshyn2021fake}. 
As of now, cyberbullying is a criminal offense in Arkansas, North Carolina, and Louisiana. Cyberbullies in North Carolina are subjected to paying a fine of up to \$1,000 or spending six months behind bars. 
In Canada, improper behavior by an individual on the Internet is considered a cyberbullying crime. Provinces in Canada have a law or legislation governing cyberbullying. The Canadian Penal Code under section 264 (2) (b) states that cyber aggression involves frequently contacting one person with another \cite{tagaymuratovna2022cyberbullying}. This offense is actionable by up to 10 years imprisonment\cite{szoka2009cyberbullying,tagaymuratovna2022cyberbullying,bocij2002online}. 
A law passed in the New South Wales administrative-territorial unit, Australia, identifies cyberbullying as a criminal offense that is applied only when cyberbullying is committed against an individual, or victim \cite{soldatova2012children}. Additionally, this law also governs all other forms of cyberbullying in our community. Like in the USA, Canada, and Australia, in Ghana,  cyberbullying on social media could lead to jailing for a minimum of 6 months and a maximum of three years in prison for cyberbullying on social media. Unlike the aforementioned countries, U.K., Switzerland, and Spain do not have cyberbullying laws.  

\section{Limitation}
The Di-CARE method was used to select relevant literature for this study. Even though the required papers were selected, there were limitations in terms of the search terms and the identified literature. Only English search terms were used. Furthermore, the list of search terms was predetermined rather than developed inductively. A second search should be conducted using the terms gathered during the literature analysis process to find additional relevant literature in the context of this literature review. Only publications of controlled quality were included in the analysis process by excluding non-peer-reviewed publications (e.g., books and white papers, surveys). Even though the books included have valuable contributions presented at conferences or published in journals, some contributions could be added to this literature review. 


\section{Conclusion}\label{conclusion}
This study used the Di-CARE method to review cyberbullying studies to identify research in cyberbullying text detection. Specifically cyberbullying forms, roles, and cyberbullying detection methods. During the review, it was identified that most researchers detected cyberbullying using machine learning algorithms, natural language processing, and deep learning. Additionally, this paper highlighted the dataset challenge in cyberbullying detection research and highlighted the impacts of law or legislation on cyberbullying across different continents.


\bibliographystyle{IEEEtran}
\bibliography{references}

\end{document}